\documentclass[pra,aps,amsmath,amssymb,amsfonts,twocolumn,nofootinbib,longbibliography,floatfix]{revtex4-1}
\usepackage{amssymb}
\usepackage{bm,mathrsfs}
\usepackage{graphicx}
\usepackage{epsfig}
\usepackage{amsmath,bbm}
\usepackage{amsfonts,amssymb}
\usepackage{times}
\usepackage{verbatim}
\usepackage[sort&compress]{natbib}
\usepackage{amsmath}
\usepackage{bm}
\usepackage{lipsum}

\usepackage{makecell}
\usepackage{array}
\usepackage{booktabs}
\usepackage{multirow}
\usepackage[colorlinks,breaklinks,linkcolor=blue,anchorcolor=blue,citecolor=blue,urlcolor=blue]{hyperref}
\begin{document}
\title {Optimal charging of open spin-chain quantum batteries via homodyne-based feedback control}
\author{Y. Yao}
\affiliation{Center for Quantum Sciences and School of Physics, Northeast Normal University, Changchun 130024,  China}
\affiliation{Center for Advanced Optoelectronic Functional Materials Research, and Key Laboratory for UV Light-Emitting Materials and Technology
of Ministry of Education, Northeast Normal University, Changchun 130024, China}

\author{X. Q. Shao}
\email{shaoxq644@nenu.edu.cn}
\affiliation{Center for Quantum Sciences and School of Physics, Northeast Normal University, Changchun 130024, China}
\affiliation{Center for Advanced Optoelectronic Functional Materials Research, and Key Laboratory for UV Light-Emitting Materials and Technology
of Ministry of Education, Northeast Normal University, Changchun 130024, China}

\begin{abstract}
We study the problem of charging a dissipative one-dimensional $XXX$ spin-chain quantum battery using local magnetic fields in the presence of spin decay. The introduction of quantum feedback control based on homodyne measurement contributes to improve various performance of the quantum battery, such as energy storage, ergotropy, and effective space utilization rate.
For the zero temperature environment, there is a set of optimal parameters to ensure that the spin-chain quantum battery can be fully charged and the energy stored in the battery can be fully extracted under the perfect measurement condition, which is found through the analytical calculation of a simple two-site spin-chain quantum battery and further verified by numerical simulation of a four-site spin-chain counterpart.
For completeness, the adverse effects of imperfect measurement, anisotropic parameter, and finite temperature on the charging process of the quantum battery are also considered.
 \end{abstract}

\maketitle
\section{Introduction}

The quantum batteries have become an important topic concerned by researchers due to the requirement for  the  efficient and miniaturized energy storage devices. Initially, the concept of quantum batteries proposed by Alicke and Fannes \cite{1} is defined as a small quantum mechanical system for temporarily storing energy. Subsequently, a series of quantum battery models have been proposed in closed systems. Simply put, the charging process can be achieved by coupling the battery to the charger which can be regarded as both energy provider and energy mediator. For simple models, such as a two-level system quantum battery or a quantum harmonic oscillator quantum battery \cite{TWO1,TWO2,TWO3,TWO4,TWO5,TWO6}, the batteries of these can achieve complete charging directly, but the stored energy has oscillation behavior because the charging process is unitary. Once the optimal charging time is missed, the stored energy of the battery is reduced obviously. To solve this problem, some researchers proposed adiabatic charging schemes to stabilize the energy storage of battery \cite{three1,three2,stable}.
In addition to the single-body quantum battery model, there are also many-body quantum battery models \cite{many2,many3,many5,many6,many7,many8,many11,SYK,TC1,Dicke1,Dicke2,Dicke3,Random,spin1,spin2,spin3,spin4,Spin5,Spin6,spin7}, including Sachdev-Ye-Kitaev batteries \cite{SYK}, Tavis-Cummings quantum battery \cite{TC1}, Dicke quantum battery \cite{Dicke1,Dicke2,Dicke3}, Random quantum batteries \cite{Random}, and spin-chain quantum battery \cite{spin1,spin2,spin3,spin4,Spin5,Spin6,spin7,spin77} and so on. These many-body quantum batteries exhibit remarkable charging speed compared with single-body quantum battery \cite{many1,many4,many9}.

As a matter of fact, the quantum batteries cannot be completely isolated from the environment. Therefore, it is necessary to investigate how to stabilize the stored energy and resist energy leakage caused by decoherence when the quantum battery is immersed in the environment
\cite{Open1,Open2,Open3,Open4,Open5,Open6,Open7,Open8,Open9,Open10,Open11,Open12,spin9,spin8,Open15,Open16,Open17,Open18,Open19,Open20,Open21}.
In 2020, Gherardini {\it et al.} \cite{Open9} put forward a stable charging scheme based on continuous measurement to compensate the entropy increase and keep the open quantum battery in the highest entropy state.
 Later, Quach and Munro \cite{Open11} introduced an open quantum battery model by using dark states to achieve superextensive capacity and power density.
 In general, quantum batteries can hardly be fully charged in the presence of environmental noise.
 Therefore how to improve the energy storage of quantum batteries in specific systems is still a challenging problem.

At the same time, we should not only pay attention to the stable and effective charging process of the battery, but also consider the maximum capacity of the battery itself, so as to provide sufficient energy for other equipment. Recently, the model of $N$-spin chain quantum battery interacting with environmental noise is constructed in Ref. \cite{spin9}, and the maximum energy of a quantum battery driven by a coherent cavity driving field or a heat reservoir is investigated. The results show that the nearest-neighbor hopping interaction enhances the energy storage and ergotropy of quantum batteries. Then, Ghosh {\it et al.} advanced an open spin-chain battery charging scheme \cite{spin8}, where they considered  each spin was connected to two local bosonic reservoirs. During the charging process, the energy dissipation channel is closed and the energy absorption channel is opened.
Their research shows that when the quantum battery is affected by Markovian noise, the storage speed of the quantum battery in the transient region is faster than that without noise, and the maximum recoverable work quantified by ergotropy is also higher.
 It is worth noting that the above results only hold when the ratio of dephasing noise rate to energy absorption rate is within a certain range, and the advantage of noise disappears when the system reaches the steady state.

In order to charge the quantum battery stably and effectively, we propose a dissipative protocol based on homodyne feedback, which is composed of $XXX$ spin chain with open boundary condition.
The energy injection into the battery is performed by applying a local external magnetic field to each spin. During the charging process, photons of spin radiation induced by the environment are collected, so that the homodyne current is generated after detection, and then the strength of local charging field becomes related to the photocurrent signal. By adjusting the feedback strength and the feedback direction, the $XXX$ spin-chain quantum battery can be fully charged and stabilized. Therefore, the energy stored by the battery in steady state is more favorable than the Markovian scheme of Ref. \cite{spin8}. The advantage of the homodyne-based feedback control mechanism is its simplicity in practical applications because it avoids the challenge of real-time state estimation required by Bayesian or state-based feedback \cite{youshi1,bijiao}. The relevant feedback theory has been widely used in various fields \cite{lilun1,lilun2,lilun3,lilun4,lilun5,lilun6} and guided by it some experimental results have been obtained \cite{shiyan1,shiyan2,shiyan3}.

The remainder of the paper is organized as follows. In Sec.~\ref{II}, we first analyze the stored energy and the maximum energy extracted from the spin-chain quantum battery (ergotropy) under feedback control, and then introduce the concept of effective space utilization rate. In Sec.~\ref{III}, we successively obtain the analytical solution of the charging process for $XXX$ spin-chain quantum battery consisting of two spins, and numerically verify the case of multiple spins.
In addition, the stochastic dynamics of the battery charging process, the effects of spin-spin interaction, the imperfect measurement, anisotropic parameter, and finite temperature on the battery charging process are also discussed in detail. Finally, we give a summary of the present protocol in Sec.~\ref{IV}.

\section{DISSIPATIVE  Spin-chain QUANTUM BATTERY}\label{II}
\begin{figure}
\centering\scalebox{0.5}{\includegraphics{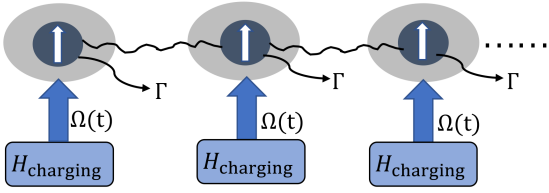}}
\caption{\label{Q1} The spin-chain quantum battery. Each spin is coupled with an independent reservoir (light grey ellipse) and has a spontaneous emission rate  $\Gamma$. Initially, the battery is prepared in its ground state and each spin is subjected to a homodyne measurement. After the detection, a corresponding homodyne photocurrent is produced. Then the feedback control is activated by applying local magnetic fields depending on the measured results.
 }
\end{figure}

\subsection{Master equation of the system}
We model the quantum battery as a one-dimensional Heisenberg spin chain consisting of $N$ spins on the lattice, each spin interacts with a local zero-temperature reservoir, as shown in Fig.~\ref{Q1}. In the absence of charging, the static Hamiltonian of the battery can be described as ($\hbar=1$)
\begin{eqnarray}\label{Hspin}
H_{B}&=&\frac{h}{2}\sum\limits_{j=1}^{N}\sigma_{j}^{z}+\sum\limits_{j=1}^{N-1}J[(1+\gamma)\sigma_{j}^{x}\sigma_{j+1}^{x}+(1-\gamma)\sigma_{j}^{y}\sigma_{j+1}^{y}\nonumber\\&&+\Delta\sigma_{j}^{z}\sigma_{j+1}^{z}],
\end{eqnarray}
where $h$ represents the strength of external magnetic field breaking the degeneracy between $\mid\downarrow\rangle$ and $\mid\uparrow\rangle$, $J$ refers to the nearest-neighbor coupling strength between spins. $\sigma^{k}$ $(k=x,y,z)$ is the corresponding Pauli spin matrix, and $\gamma$ and $
\Delta$ are the anisotropic parameters. For convenience, we assume that all the parameters here are positive real numbers.

To inject energy into battery, the local external magnetic field needs to be applied to each spin. The form of charging Hamiltonian is chosen as:
\begin{equation}\label{HC}
 H_{\rm charging}=\sum_{j=1}^{N}\{\Omega_{j}[\sigma_{j}^{x}\sin(\alpha)+\sigma_{j}^{y}\cos(\alpha)]\},
\end{equation}
where $\alpha\in(-\pi,\pi]$ is viewed as a regulator of changing the direction of magnetic field, and $\Omega_{j}$ is the corresponding strength applied to the $j$th spin. The evolution of the system can be characterised by the following Lindblad master equation
\begin{equation}\label{rou}
\mathop{\dot{\rho}}=-i[H_{B}+H_{\rm charging},\rho]+\Gamma\sum_{j=1}^{N}\mathcal{D}[\sigma_{j}^{-}]\rho,
\end{equation}
where $\mathcal{D}[o]\bullet=o\bullet o^{\dag}-({o^{\dag}o\bullet+\bullet o^{\dag}o})/2$ and $\Gamma$ is the spontaneous emission rate of each spin constituting the quantum battery.

If the above model is considered in a closed system, the energy stored in the battery usually oscillates over time, and the protocol is only effective for the instantaneous charging of the quantum battery, and in actual environment, the non-zero value of $\Gamma$ further reduces the charging performance of the battery. Therefore, designing an efficient charging scheme in open quantum systems is a problem worthy of consideration. From this perspective,  we introduce the quantum feedback mechanism based on homodyne measurement into the charging protocol of quantum battery.

In the  process of feedback control, the homodyne measurement of each spin is performed, and the resulting photocurrent $J^{\rm hom}_{j}$ can be approximated as signal plus Gaussian white noise \cite{hom1,hom2,weiner1,weiner2}, i.e., (please see appendix.~\ref{A})
\begin{equation}\label{Jom}
J^{\rm hom}_{j}(t)=\langle\sigma_{j}^{x}\rangle+\frac{\xi(t)}{\sqrt{\eta\Gamma}},
\end{equation}
where $\eta\leq1$ is the total measurement efficiency of the detection, and $\xi(t)=dw(t)/dt$ represents Gaussian white noise with a complex Wiener increment $dw(t)$ satisfying $[dw(t)]^{2}=dt$ and an average over the noise $E[dw(t)]=0$ \cite{weiner1}. 
The local magnetic field thus becomes time-dependent and its strength depending on the measurement results can be expressed as
\begin{equation}\label{Hcha}
\Omega_{j}(t)=\Omega_{0j}+fJ^{\rm hom}_{j}(t-\tau),
\end{equation}
where $f$ is the feedback strength, and $\tau$ represents a small time delay in the feedback loop. For simplicity, the constant amplitude $\Omega_{0j}$ can be set to 0, then the corresponding feedback Hamiltonian reads

\begin{equation}\label{Hre}
H_{\rm fb}=\sum_{j=1}^{N}J^{\rm hom}_{j}(t-\tau)F_{j},
\end{equation}
where $F_{j}=f[\sigma^{x}_{j}\sin(\alpha)+\sigma^{y}_{j}\cos(\alpha)]$. Now the total conditioned equation including feedback obeys the following rule \cite{weiner2}
\begin{equation}\label{zhuangtai}
\rho_{J}(t+dt)=\sum_{j=1}^{N}e^{\mathcal{K}_{j}J^{\rm hom}_{j}(t-\tau)dt}(\rho_{J}(t)+\mathcal{A}[\rho_{J}(t)]),
\end{equation}
where $\mathcal{K}_{j}$ is a Liouville superoperator satisfying $\mathcal{K}_{j}\rho=-i[F_{j},\rho]$.
$\mathcal{A}[\rho_{J}(t)]=\Gamma\mathcal{D}[\sigma^{-}_{j}]\rho_{J}(t)dt+\mathcal{H}[-iH_{B}]\rho_{J}(t)dt+\sqrt{\eta\Gamma}d\omega(t)\mathcal{H}[\sigma^{-}_{j}]\rho_{J}(t)
$ with $\mathcal{H}$ being a nonlinear superoperator defined as $\mathcal{H}[o]\bullet=o\bullet+\bullet o^{\dag}-{\rm Tr}\{\bullet(o+o^{\dag})\}\bullet$. 
Under the Markovian limit of $\tau\rightarrow0$, Eq.~(\ref{zhuangtai}) can be expanded to second order in $\mathcal{K}_{j}$ while retaining the first order of $dt$. Then, by applying the rules of Ito calculus (i.e., $[dw(t)]^{2}=dt$ ), the stochastic master equation describing the charging process of the spin-chain quantum battery is
\begin{eqnarray}\label{HF}
\mathop{\dot{\rho}_{J}}&=&\sum_{j=1}^{N}\{\mathcal{K}_{j}(\sigma^{-}_{j}\rho_{J}+\rho_{J}\sigma^{+}_{j})+\frac{1}{2\eta\Gamma}\mathcal{K}_{j}^{2}\rho_{J}
+\sqrt{\eta\Gamma}\xi(t)\nonumber\\&&\times[\mathcal{H}[\sigma^{-}_{j}]+(\eta\Gamma)^{-1}\mathcal{K}_{j}]\rho_{J}\}+\mathcal{L}\rho_{J},
\end{eqnarray}
where $\mathcal{L}\rho_{J}=-i[H_{B},\rho_{J}]+\Gamma\sum_{j=1}^{N}\mathcal{D}[\sigma^{-}_{j}]\rho_{J}$. Taking the ensemble average of Eq.~(\ref{HF}), we have
\begin{eqnarray}\label{M}
\mathop{\dot{\rho}}&=&-i[H_{B},\rho]+\Gamma\sum_{j=1}^{N}\mathcal{D}[\sigma^{-}_{j}]\rho\nonumber\\&&-i\sum_{j=1}^{N}\{[F_{j},\sigma_{j}^{-}\rho+\rho\sigma_{j}^{+}]+\frac{1}{2\eta\Gamma}[F_{j},-i[F_{j},\rho]]\},
\end{eqnarray}
where the first two terms describe the dynamical evolution caused by the internal Hamiltonian of battery and the spontaneous decay, respectively. The last two terms describe the charging coherence introduced by the feedback operation and the measurement noise fed back into the system. The central idea is to  use continuous measurement records to control the dynamics of system.

\subsection{The relevant performance parameters of the battery}
The stored energy of the quantum battery at an arbitrary time $t$ is defined as
\begin{equation}\label{energy}
\Delta E(t)={\rm Tr}[H_{B}\rho_{B}(t)]-{\rm Tr}[H_{B}\rho_{B}(0)],
\end{equation}
where $\rho_{B}(t)$ is the density matrix of the battery and $\rho_{B}(0)$ is the initial state.
In order to describe the maximum energy that can be extracted from the battery, we need to calculate the corresponding ergotropy \cite{tiqu}
\begin{equation}\label{ergotropy}
\mathcal{E}(t)={\rm Tr}[H_{B}\rho_{B}(t)]-\mathop{\rm min}\limits_{U}{\rm Tr}[U\rho_{B}(t)U^{\dag}H_{B}],
\end{equation}
the second term in the r.h.s. of Eq.~(\ref{ergotropy}) indicates the minimum energy value which cannot be extracted from the battery by executing all unitaries $U$ on the system.

If Hamiltonian $H$ and density matrix $\rho$ of a system are written orderly as $H=\sum_{i}\varepsilon_{i}|\varepsilon_{i}\rangle\langle \varepsilon_{i}|$ ($\varepsilon_{1}<\varepsilon_{2}\cdots< \varepsilon_{N}$), and $\rho=\sum_{k}r_{k}|r_{k}\rangle\langle r_{k}|$ ($r_{1}>r_{2}\cdots> r_{N}$), respectively, the corresponding passive state \cite{tiqu, passive} of the system can be expressed as
\begin{equation}\label{passive}
\sigma=\sum_{j}r_{j}|\varepsilon_{j}\rangle\langle \varepsilon_{j}|.
\end{equation}
Thus the ergotropy can be explicitly rewritten as
\begin{eqnarray}\label{ergotropy1}
\mathcal{E}(t)&=&{\rm Tr}[H_{B}\rho_{B}(t)]-{\rm Tr}[H_{B}\sigma]\nonumber\\
         &=&{\rm Tr}[H_{B}\rho_{B}(t)]-\sum_{j}r_{j}\varepsilon_{j}.
\end{eqnarray}

It is well known that the energy spectrum structure of a spin chain is closely related to the nearest-neighbor coupling strength between spins, thus it is not comprehensive to measure the performance of the battery only by the stored energy of the battery and the ergotropy. Here, we introduce the effective space utilization rate of the battery, i.e. the ratio of the total battery space occupied by the stored energy of the battery, which is also an important parameter for benchmarking the performance of quantum battery with internal interaction,
\begin{equation}\label{R}
R(t)=\frac{\Delta E(t)}{C_{\rm max}},
\end{equation}
where $C_{\rm max}=E_{\rm max}-E_{\rm min}$ is the maximum capacity of the battery, and $E_{\max}$ and $E_{\min}$ are the highest energy and the lowest energy in the energy spectrum of $H_{B}$, respectively. Undoubtedly, $R=1$ indicates that the battery can be fully charged.

\section{the OPTIMAL FEEDBACK CONDITION}\label{III}

\subsection{The charging of two-spin quantum battery under homodyne-based feedback control}
For convenience, we mainly take the $XXX$ spin-chain quantum battery ($\Delta=1$ and $\gamma=0$) consisting of two spins as an toy model to study the charging process. This simple model can not only reflect the characteristics obviously different from the single-body quantum battery, but also help us  determine the optimal parameter range when extending to a many-body spin quantum battery. According to
Eq.~(\ref{Hspin}), we can get the following eigenstates
\begin{subequations}\label{c}
\begin{align}
|E_{a}\rangle &=\mid\downarrow\downarrow\rangle,\label{ca}\\
|E_{b}\rangle &=\frac{1}{\sqrt{2}}(\mid\uparrow\downarrow\rangle-\mid\downarrow\uparrow\rangle), \label{cb}\\
|E_{c}\rangle &=\frac{1}{\sqrt{2}}(\mid\uparrow\downarrow\rangle+\mid\downarrow\uparrow\rangle), \label{cc}\\
|E_{d}\rangle &=\mid\uparrow\uparrow\rangle,\label{cd}
\end{align}
\end{subequations}
with the corresponding eigenvalues $E_{a}=-h+J$, $E_{b}=-3J$, $E_{c}=J$, and $E_{d}=h+J$, respectively. It is easy to check that the order of eigenvalues is $E_{a}<E_{b}<E_{c}<E_{d}$ when $J<h/4$, and the order of eigenvalues becomes $E_{b}<E_{a}<E_{c}<E_{d}$ once $J>h/4$.  The above results show that the ground state  of the system is related to the coupling strength $J$ between spins. But the highest energy state of the battery is always $|E_{d}\rangle =\mid\uparrow\uparrow\rangle$ regardless of $J$.

The steady-state solution of  Eq.~(\ref{M}) (see appendix.~\ref{B}) is
\begin{subequations}\label{w}
\begin{align}
\rho_{11}(\infty)&=\frac{\chi^{4}}{(2\chi^{2}+2\chi\eta\cos\alpha+\eta)^2}, \label{wa}\\
\rho_{22}(\infty)&=\frac{\chi^{2}(\chi^{2}+2\chi\eta\cos{\alpha}+\eta)}{(2\chi^{2}+2\chi\eta\cos{\alpha}+\eta)^{2}}, \label{wb}\\
\rho_{33}(\infty)&=\rho_{22}(\infty), \label{wc}\\
\rho_{44}(\infty)&=\frac{(\chi^{2}+2\chi\eta\cos{\alpha}+\eta)^{2}}{(2\chi^{2}+2\chi\eta\cos{\alpha}+\eta)^{2}}, \label{wd}
\end{align}
\end{subequations}
where we have assumed $\chi=f/\Gamma$ for simplicity. The optimal feedback condition can be determined through  $\partial\rho_{11}(\infty)/\partial\chi=2(\chi^{2}\eta\cos\alpha+\chi\eta)/(2\chi^{2}+2\chi\eta\cos\alpha+\eta)^{2}=0$. By selecting $\chi=-1/\cos\alpha$, the maximum value of $\rho_{11}(\infty)$ is obtained as
\begin{equation}\label{rou11}
\rho_{11}^{\rm max}(\infty)=\frac{1}{(2-\eta\cos^{2}{\alpha})^{2}}.
\end{equation}
Under the perfect measurement condition $\eta=1$, the optimal value of $\rho_{11}^{\rm max}(\infty)$ is 1 when $\alpha=0$ or $\pi$, which means the battery can be fully charged [$R(\infty)$=1] from its ground state by applying $y$-direction feedback to the battery, even in the presence of dissipation. Thus, $\alpha=0$ $(\pi)$ and $\chi=\mp1$ are viewed as the optimal feedback parameters of the $XXX$ spin-chain quantum battery.
In what follows,
we stipulate that the $y$-direction feedback refers to the case of $\alpha =\pi$, unless otherwise specified.
 According to Eqs.~(\ref{energy}) and (\ref{ergotropy1}), the corresponding ratio of ergotropy to the stored energy of the battery in steady state is computed to be $\mathcal{E}(\infty)/\Delta E(\infty)=1$, whenever the ground state of the $XXX$ model is $\mid\downarrow\downarrow\rangle$ ($J<h/4$) or $(\mid\uparrow\downarrow\rangle-\mid\downarrow\uparrow\rangle)/\sqrt{2}$ ($J>h/4$), which means the stored energy in the battery can be fully extracted.

\subsection{The charging process of the battery made up with four spins}\label{IIIB}
In the above study, we mainly take the spin-chain quantum battery composed of two spins as an example to analyze the charging performance of the battery. In order to verify the similar conclusion is also applicable to the battery with more spins, we now numerically discuss the case of spin-chain quantum battery composed of four spins. Considering more spins will undoubtedly increase our computational time, but will not change the conclusion. In the numerical simulation, we set the ground state of the battery as the initial state, and the measurement efficiency $\eta$ is 1.

\begin{figure}
\centering\scalebox{0.115}{\includegraphics{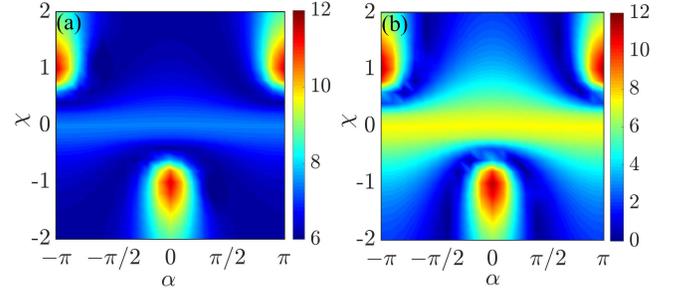}}
\caption{\label{PP} (a) The stored energy $\Delta E(\infty)/h$ of the battery at steady state is a function of $\alpha$ and $\chi$. (b) The maximum energy $\mathcal{E}(\infty)/h$ that can be extracted (ergotropy) from the battery, other parameters are $\eta=1$ and $J=h$.}
\end{figure}

In order to determine the optimal feedback parameters, the stored energy and the ergotropy of the battery in steady state are displayed as functions of $\alpha$ and $\chi$ in Fig.~\ref{PP}, governed by Eq.~(\ref{M}), and other parameters are $\eta=1$ and $J=h$.
It clearly shows that there are three equivalent optimal parameters ($\alpha$, $\chi$) in the corresponding parameter range, which is consistent with the case of two spins. (Please see  appendix.~\ref{Bbb} for more spins). For convenience, we choose the optimal feedback condition ($\alpha=\pi$, $\chi=1$) for the following discussion.

\begin{figure}
\centering\scalebox{0.23}{\includegraphics{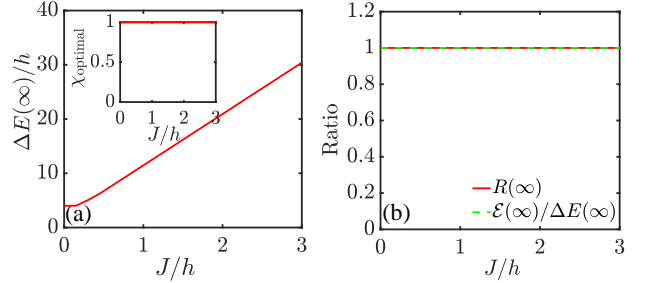}}
\caption{\label{P6} The $y$-direction feedback control is applied to the system, i.e., $\alpha=\pi$. (a) The variation of optimal stored energy $\Delta E(\infty)/h$ of the battery at steady state with $J/h$. The inset describes the optimal feedback parameter $\chi$ at different $J$. (b) The variations of $R(\infty)$ and $\mathcal{E}(\infty)/\Delta E(\infty)$ of the battery in the steady state with the coupling strength $J$. The measurement efficiency is given as $\eta=1$.
 }
\end{figure}

Fig.~\ref{P6}(a) shows that the optimal stored energy of the battery in the steady state increases with the increase of coupling strength $J/h$ between spins, which means the interaction between spins expands the capacity of the battery itself. From the inset, we see the feedback strength required for the battery to achieve the optimal energy storage does not change with $J$. In order to  measure the battery performance under feedback mechanism more comprehensively, we characterize the variation of the space utilization rate [$R(\infty)$] and the ratio of the ergotropy to the stored energy [$\mathcal{E}(\infty)/\Delta E(\infty)$] of the battery with the coupling strength between spins in Fig.~\ref{P6}(b). It indicates that the coupling strength between spins is not the factor affecting the ability of energy storage and energy extraction, because the highest energy state of the $XXX$ spin-chain battery is always the state all spins pointing upward, irrespective to the coupling strength between the spins.  After the above analysis of four spins, we qualitatively obtain the conclusion similar to the case of two spins. It clearly proves that our scheme is still effective for quantum battery composed of multiple spins.

\subsection{Random dynamics, imperfect measurement, and different initial state}
\begin{figure}
\centering\scalebox{0.4}{\includegraphics{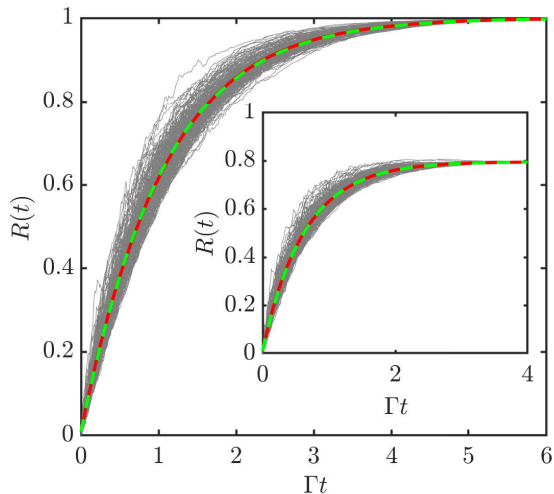}}
\caption{\label{Q4} Stochastic charging dynamics of the battery under the optimal feedback condition. In the inset, the measuring efficiency is set to $\eta=0.8$. The shallow grey lines represent the random 200 trajectories of battery storage energy obtained by  Eq.~(\ref{HF}), the red solid line represents the average value of 200 energy trajectories, and green dashed line represents the ensemble average energy obtained by Eq.~(\ref{M}).
 }
\end{figure}

In order to show the stochastic nature of the measurement and feedback processes, in Fig.~\ref{Q4}, we numerically simulate 200 random trajectories of the energy stored in the $XXX$ spin-chain battery varying with $\Gamma t$ based on the random master equation Eq.~(\ref{HF}). Each trajectory represents the result of a single operation (see the light grey line). Obviously, the energy of different trajectories has dispersion behavior in the charging process. But, when the system reaches the steady state, the feedback has a strong inhibitory effect on the fluctuation of battery energy. In addition, the average of 200 energy trajectories (the red solid line) is coincide with the energy of the ensemble average obtained through the master equation Eq.~(\ref{M}) (the green dashed line). The above results show that the battery can be fully charged not only in the overall average sense, but also in the single-trajectory. And the inset shows that the battery can be stably and effectively charged  even if the measurement efficiency is not perfect.
\begin{figure}
\centering\scalebox{0.23}{\includegraphics{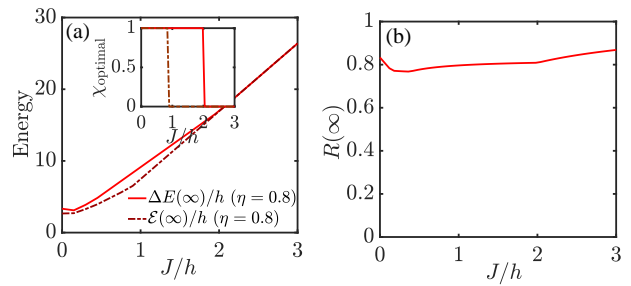}}
\caption{\label{P7} (a) The stored energy $\Delta E(\infty)/h$ (solid line) and the ergotropy (dash-dotted line) of the spin-chain quantum battery with different coupling strengths $J/h$. The inset shows the optimal feedback parameter $\chi$ required by the stored energy (solid line) and ergotropy (dash-dotted line). (b) The effective utilization rate $R(\infty)$ of the battery as a function of $J/h$ under the optimal feedback strength. The measurement efficiency is  $\eta=0.8$.}
\end{figure}

Fig.~\ref{P7}(a) describes the optimal stored energy (solid line) and ergotropy (dash-dotted line) of the battery in the steady state with imperfect measurement efficiency $\eta=0.8$. The corresponding optimal feedback strengths with different $J$ values are shown in the inset. We can clearly see that the maximum energy extracted from the battery is lower than the optimal stored energy of the battery. When the spin-spin interaction strength increases to a certain value $J_{c}$ ($J_{c}=2h$ in terms of energy storage and $J_{c}=0.8763h$ in terms of energy extraction, please see appendix~\ref{C} for details), the  required optimal feedback intensity should be zero, which means that the feedback control is invalid in this case.
Fig.~\ref{P7}(b) describes the variation of the effective space utilization rate $R(\infty)$ of the battery with the $J$ at the optimal feedback strength (the inset of Fig.~\ref{P7}(a)). The results reflect that the imperfect measurement weakens the role of feedback control, and reduces the performance of the battery.

\begin{figure}
\centering\scalebox{0.4}{\includegraphics{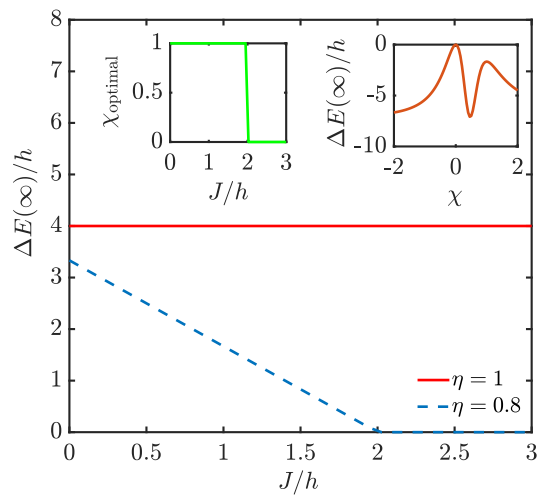}}
\caption{\label{P8} The stored energy $\Delta E(\infty)/h$ of the battery in steady state as a function of $J/h$ under different measurement efficiency $\eta$. The initial state of the battery is spin-down state and the feedback parameter $\chi$ at different $J$ values is optimal.}
\end{figure}

So far, we have studied the charging performance of spin-chain quantum battery using the ground state as the initial state. Next, we mainly study the charging process of the battery when all the spin are initialized downward.  Fig.~\ref{P8} reveals that the stored energy of the battery is independent of $J$ under the condition of perfect measurement (red solid line), because the stored energy of the battery during the whole charging process is $\Delta E(\infty)=(3J+2h)-(-2h+3J)=4h$. Nevertheless, when the measurement efficiency is imperfect ($\eta=0.8$), the stored energy of the battery
at steady state is greatly affected (blue dash line). We can see again that the  nearest-neighbor hopping between spins has a critical value that renders the feedback control ineffective, as shown in the inset on the left.  To better understand this phenomenon, we use $J/h=3$ to give the energy storage curve of the battery with the change of feedback strength in the right inset. The result manifests that the stored energy of the battery becomes negative when the feedback control is switched on, indicating that the battery has evolved into a lower energy state than the initial state, where the battery is in the discharging process.

\subsection{Performance of $XY$ spin-chain quantum battery and $XYZ$ spin-chain quantum battery}
The effect of anisotropic parameter ($\gamma$)  on the charging process of the battery can be characterized by analyzing the related properties of
 the $XY$ spin-chain quantum battery ($\Delta=0$, $0<\gamma<1$) and $XYZ$ spin-chain quantum battery ($\Delta=1$, $0<\gamma<1$). In this part, we assume the measurement efficiency is $\eta=1$, and the feedback parameter $\chi$ takes the optimal value under their respective $\gamma$.

\begin{figure}
\centering\scalebox{0.43}{\includegraphics{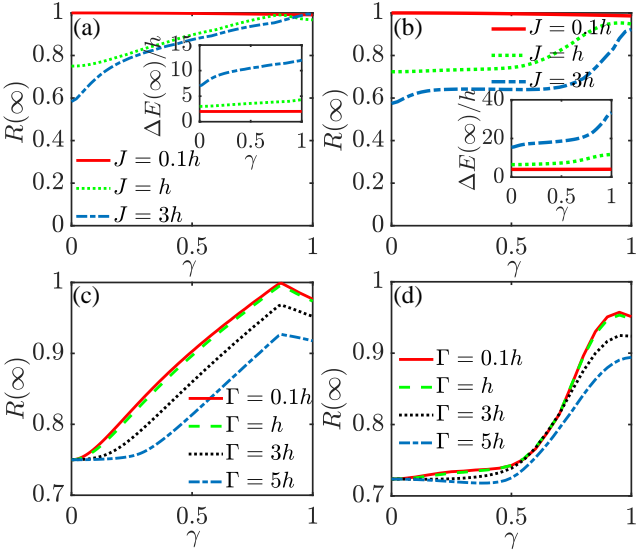}}
\caption{\label{P11} The effective space utilization rate $R(\infty)$ in steady state of the $XY$ spin-chain quantum battery ($\Delta=0$) with the spin number of $N=2$ (a) and $N=4$ (b) as a function of the anisotropic parameter $\gamma$ under different coupling strengths $J$. The insets in (a) and (b) describe the variations of the stored energy $\Delta E(\infty)/h$ of the battery with $\gamma$ in steady state. The feedback strengths at different $\gamma$ values are optimal.
Other parameters are $\eta=1$, $\alpha=\pi$ and $\Gamma=h$. (c) and (d) show the variations of the effective space utilization with $\gamma$ of the battery with spin number of $N=2$ and $N=4 $ under different spontaneous emission rates $\Gamma$ with $J=h$, respectively.
 }
\end{figure}

Figs.~\ref{P11}(a) and \ref{P11}(b) respectively describe the influence of the anisotropic parameter $\gamma$ on the effective space
utilization rate $R(\infty)$ and the optimal stored energy $\Delta E(\infty)/h$ of the $XY$ spin-chain battery when the spin numbers of $N=2$ and $N=4$. Different curves correspond to different spin-spin coupling strengths, where the spontaneous emission rate is set to $\Gamma=h$.  The results in Fig.~\ref{P11}(a) and the inset show that for a weak $J = 0.1h$ the anisotropic parameter $\gamma$ almost does not affect the effective space utilization rate $R(\infty)$  and the optimal stored energy $\Delta E(\infty)$. However, for a relatively strong $J=3h$, $R(\infty)$ and $\Delta E(\infty)$ increase with $\gamma$ significantly. When the number of particles increases to $N=4$, we can obtain a conclusion similar to that of $N=2$, as shown in Fig.~\ref{P11}(b). The above results indicate that under the strong interaction between spins, the increase of anisotropic parameter $\gamma$ plays a positive role in improving the performance of the $XY$ spin-chain quantum battery.
Figs.~\ref{P11}(c) and \ref{P11}(d) respectively describe that the effective space utilization rates of the battery with $N=2$ and $N=4$ decrease with the increase of dissipation rate under a given spin-spin coupling strength ($J=h$), which is completely different from the $XXX$ model mentioned above. Nevertheless, a relatively weak dissipation rate ranging from $0.1h$ to $h$, can still be regarded as an independent variable.
It is worth noting that the specific values of $\gamma=0$ and $\gamma=1$ correspond to the typical transverse $XX$ model and transverse Ising model respectively.

\begin{figure}
\centering\scalebox{0.43}{\includegraphics{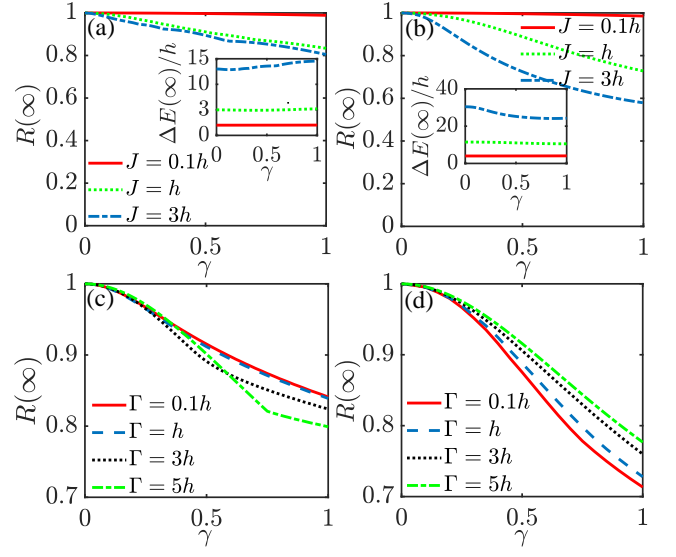}}
\caption{\label{P12}(a) The effective space utilization rate $R(\infty)$ of the $XYZ$ spin-chain quantum battery ($\Delta=1$) in steady state as a function of the anisotropic parameter $\gamma$ for $N=2$. The inset describes the change of the effective space utilization rate $R$ of the battery with $\gamma$ in steady state. (b) The same as (a) but the spin number is $N=4$. Other conditions are the same with the Fig.~\ref{P11}. (c) and (d) show the variations of the effective space utilization with $\gamma$ of the battery with spin number of $N=2$ and $N=4 $ under different spontaneous emission rates $\Gamma$ with $J=h$, respectively.
 }
\end{figure}

The influence of anisotropic parameter $\gamma$ on the effective space utilization rate and optimal energy storage of $XYZ$ spin-chain quantum batteries with $N = 2$ and $N = 4$ is illustrated in Figs.~\ref{P12}(a) and \ref{P12}(b) with $\Gamma=h$ respectively.
It indicates that no matter $N=2$ or $N=4$, when the spin coupling strength is weak ($J=0.1h$), the anisotropic parameters still have no significant effect on the effective space utilization rate and optimal energy storage.
However, when the coupling strength between spins increases to $J=3h$,  with the increase of anisotropic parameters, the batteries with $N=2$ and $N=4$ exhibit different behaviors in the terms of optimal energy storage.
Meanwhile, we see the effective space utilization of the batteries with spins of $N=2$ and $N=4$ decreases with the increase of $\gamma$, which means that the increase in anisotropy is detrimental to the battery chargeability.
Figs.~\ref{P12}(c) and \ref{P12}(d) show the effects of different spontaneous emission rates $\Gamma$ on the effective space utilization rate of the $XYZ$ spin-chain quantum batteries with the spin number of $N=2$ and $N=4$, respectively. 
The influence of spontaneous emission rate on the current model is obviously different from that of $XY$ spin-chain model, especially shown in Figs.~\ref{P12}(d),  the effective space utilization increases with the increase of dissipation rate.
Similarly, the specific value of $\gamma=0$ corresponds to the $XXX$ model in  Sec.~\ref{IIIB}, and we have shown in the previous discussion that its performance is not affected by spin-spin interaction strength and the spontaneous emission rate.

\subsection{The effect of finite temperature}\label{E}

\begin{figure}
\centering\scalebox{0.43}{\includegraphics{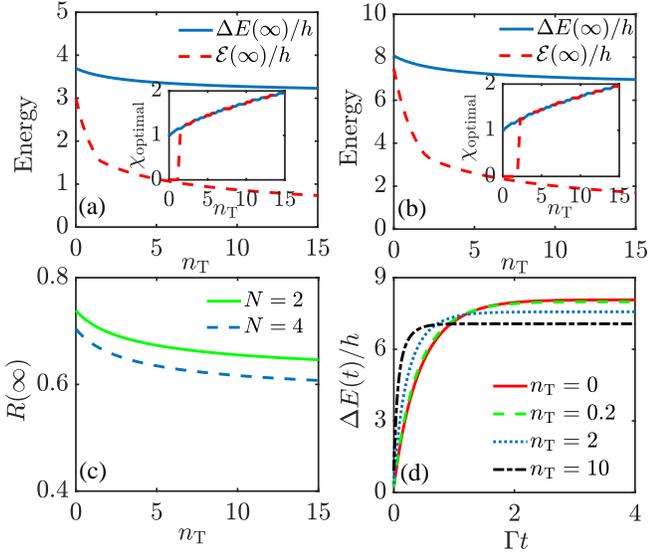}}
\caption{\label{P001}(a)-(b) describe the optimal energy storage and ergotropy of the battery at steady state when the spin numbers are $N=2$ and $N=4$, respectively. And the corresponding optimal feedback parameter $\chi$ is shown in the inset. (c) describes the variation of effective space utilization rate of the battery with the temperature. (d) shows the evolution of energy stored in the battery ($N=4$) with time $\Gamma t$ at different temperatures, where the feedback strengths are adjusted to the optimal values and the initial state of the battery is set to the corresponding ground state. Other parameters are $\eta_{d}=\eta_{c}=0.8$ and $J=h$.
 }
\end{figure}

When there is a finite temperature $T$ in the environment, the uncollected photons will enter the thermal radiation reservoir with an average occupancy of $n_{\rm T}=[\exp(\hbar\omega/k_{B}T)-1]^{-1}$. Therefore, we assume that the photons radiated by spin enter two channels, one is the thermal reservoir with finite temperature for the uncollected photons, the other is the effective zero temperature for the collected photons. Then the conditional state of the system influenced by homodyne measurement can be written as
\begin{eqnarray}\label{T1}
\dot{\rho}_{J}&=&-i[H_{B},\rho_{J}]+\sum_{j=1}^{N}\{\eta_{c}\Gamma\mathcal{D}[\sigma^{-}_{j}]\rho_{J}+\sqrt{\eta\Gamma}\xi(t)\mathcal{H}[\sigma^{-}_{j}]\rho_{J}\}\nonumber\\&&
+(1-\eta_{c})\sum_{j=1}^{N}\mathcal{L}_{j\rm T}\rho_{J},
\end{eqnarray}
where $\mathcal{L}_{j\rm T}\rho_{J}=\Gamma\{(1+n_{\rm T})\mathcal{D}[\sigma^{-}_{j}]\rho_{J}+n_{\rm T}\mathcal{D}[\sigma^{+}_{j}]\rho_{J}\}$ is the standard Lindblad dissipator of the thermal radiation reservoir \cite{Open20,weiner2}. $\eta$ is the total measurement efficiency of the detection system satisfying $\eta=\eta_{c}\eta_{d}$, which incorporates the collected photon efficiency $\eta_{c}$ and the detector efficiency $\eta_{d}$ simultaneously.

The master equation based on the homodyne feedback control should be rewritten as
\begin{eqnarray}\label{T2}
\mathop{\dot{\rho}}&=&-i[H_{B},\rho]+\eta_{c}\Gamma\sum_{j=1}^{N}\mathcal{D}[\sigma^{-}_{j}]\rho+(1-\eta_{c})\sum_{j=1}^{N}\mathcal{L}_{j\rm T}\rho\nonumber\\&&
-i\sum_{j=1}^{N}\{[F_{j},\sigma_{j}^{-}\rho+\rho\sigma_{j}^{+}]+\frac{1}{2\eta\Gamma}[F_{j},-i[F_{j},\rho]]\}.
\end{eqnarray}

By solving the Eq.~(\ref{T2}) with $N=2$, we find that the terms in the steady state are given by
\begin{subequations}\label{T3}
\begin{align}
\rho_{11}(\infty)&=\frac{(f^{2}+n_{\rm T}\Gamma^{2}\eta-n_{\rm T}\Gamma^{2}\eta\eta_{c})^{2}}{[2f^{2}-2f\Gamma\eta+(1+2n_{\rm T})\Gamma^{2}\eta-2n_{\rm T}\Gamma^{2}\eta\eta_{c}]^{2}},\label{T3a}\\
\rho_{22}(\infty)&=-\frac{\Gamma^{2}(-2f+\Gamma)^{2}\eta^{2}}{4[2f^{2}-2f\Gamma\eta+(1+2n_{\rm T})\Gamma^{2}\eta-2n_{\rm T}\Gamma^{2}\eta\eta_{c}]^{2}}\nonumber\\& \quad+\frac{1}{4},\label{T3b}\\
\rho_{33}(\infty)&=\rho_{22}(\infty),\label{T3c}\\
\rho_{44}(\infty)&=\frac{[f^{2}-2f\Gamma\eta+(1+n_{\rm T})\Gamma^{2}\eta-n_{\rm T}\Gamma^{2}\eta\eta_{c}]^{2}}{[2f^{2}-2f\Gamma\eta+(1+2n_{\rm T})\Gamma^{2}\eta-2n_{\rm T}\Gamma^{2}\eta\eta_{c}]^{2}}.\label{T3d}
\end{align}
\end{subequations}
For a given temperature and collection efficiency, the optimal feedback strength $f$ for the stored energy of the battery can be determined as
\begin{equation}\label{T4}
f/\Gamma=\frac{1}{2}[1+\sqrt{1+4n_{\rm T}\eta(1-\eta_{c})}],
\end{equation}
and the corresponding maximum value of the $\rho_{11}(\infty)$ is obtained as follows
\begin{equation}\label{T5}
\rho_{11}^{\rm max}(\infty)=\frac{1}{4}[1+\frac{\eta\sqrt{1+\mu}}{1+\mu+(1-\eta)\sqrt{1+\mu}}]^{2},
\end{equation}
where $\mu=4n_{\rm T}\eta(1-\eta_{c})$. The existence of finite temperature increases the required optimal feedback strength $f$ and reduces the population of the system in $\rho_{11}(\infty)$, and cause the battery to not be fully charged.

In Fig.~\ref{P001}, we set the coupling strength between spins as $J=h$, and the total feedback efficiency is $\eta=0.64$. Whether the spin chain quantum battery is composed of two spins (Fig.~\ref{P001} (a)) or four spins (Fig.~\ref{P001} (b)), it can be clearly seen that the  ergotropy and stored energy of battery in the steady state decrease monotonously with the increase of temperature. Moreover, the finite temperature increases the required optimal feedback strength as shown by the corresponding inset, we once again observe that the feedback control will fails in term of the ergotropy when the temperature is relatively low, which is similar to the behavior in Fig.~\ref{P7} (a).  Fig.~\ref{P001} (c) depicts the change of the effective space utilization rate of the battery with the temperature, and the results show that $R(\infty)$ decreases with the increase of temperature. Finally, the energy stored of the battery at different temperatures is plotted as a function of $\Gamma t$ in the Fig.~\ref{P001} (d), where the feedback intensity is set to the corresponding optimal value. It indicates that the time to reach the steady state of the system is shortened with the increase of temperature. Although the finite temperature is not conducive to the battery energy storage, it can still be maintained at a high value when the temperature is relatively low ($n_{\rm T}\in[0,0.2])$.

\section{Summary}\label{IV}
In summary, we have constructed a dissipative $XXX$ spin-chain quantum battery model based on homodyne measurement. Its core idea is to use continuous measurement records to control system dynamics. The analytical form of $XXX$ spin chain model of two spins show that the battery can be charged stably and fully under the optimal feedback condition.
We also extend the model to the case of four spins, and qualitatively get conclusions similar to that of two-spin model. The interaction between spins expands the capacity of the $XXX$ spin-chain battery itself, but does not affect the degree of storage and extraction of battery energy. 
The stochastic dynamics of the battery charging process reflects that the present feedback mechanism is not only effective in the overall average sense, but also effective in a single trajectory. 
The imperfect measurement and the finite temperature weaken the influence of feedback control on the system, which are not conducive to the storage and extraction of battery energy. In addition, under the imperfect measurement condition, there is a critical value $J_{c}$ of the coupling strength,  and beyond this critical value, the feedback control will fail. 
For the $XY$ spin-chain quantum battery, the increase of anisotropy is beneficial to improve the effective space utilization rate. Meanwhile, the lower dissipation rate is more favorable for battery energy storage. For the $XYZ$ spin-chain quantum battery, the relatively large spontaneous emission will help to improve the effective space utilization rate. We hope that our work will provide a
valuable reference for future research on open quantum batteries.

\section*{Acknowledgements}
X. Q. Shao would like to express his
thanks to Dr. Lingzhen Guo  for his valuable discussion on the numerical simulation of homodyne-based feedback control, and Dr. Gangcheng Wang for his helpful comment on the feature of $XXX$ spin-chain model.
The anonymous reviewers are also thanked for constructive
comments that helped in improving the quality of this paper. This work is supported by the National Natural Science Foundation of China (NSFC) under Grants No. 11774047 and No. 12174048.

\appendix
\section{The relevant derivation of Eq.~(\ref{Jom})}\label{A}

In the feedback control process, the photon emitted by each spin is collected and becomes a beam with annihilation operator $\sqrt{\Gamma} \sigma^{-}_{j}$. Then the beam enters one port of a 50 : 50 beam splitter, and a strong local oscillator $\beta$ enters another port \cite{hom1,hom2,weiner1,weiner2}. After output by beam splitter, two field operators $b_{1j}$ and $b_{2j}$ are obtained in the form of
\begin{equation}\label{A1}
b_{kj}=[\sqrt{\Gamma} \sigma^{-}_{j}-(-1)^{k}\beta]/\sqrt{2},
\end{equation}
when both fields are detected, the corresponding forms of two photocurrent are
\begin{equation}\label{A2}
\overline{J_{kj}}=\langle|\beta|^{2}-(-1)^{k}(\sqrt{\Gamma}\beta\sigma^{+}_{j}+\sqrt{\Gamma}\beta^{\ast}\sigma^{-}_{j})+\Gamma\sigma^{+}_{j}\sigma^{-}_{j}\rangle/2,
\end{equation}
where $\sqrt{\Gamma}\beta\sigma^{+}_{j}+\sqrt{\Gamma}\beta^{\ast}\sigma^{-}_{j}$ represents the interference of the system and the local oscillator. The above equation refers to the average photocurrent. 
However, instantaneous photocurrent rather than average photocurrent is recorded in practice.
Due to the randomness in the quantum measurement process, the photocurrent will change randomly.  In the ideal case of homodyne detection, namely $|\beta|^{2}\gg\Gamma$, the rate of  photodetections tends to be infinite, and the corresponding form of photocounts can be approximated as signal plus Gaussian white noise.
\begin{equation}\label{A3}
J^{\rm {hom}}_{j}(t)=\frac{J_{1j}(t)-J_{2j}(t)}{|\beta|}=\sqrt{\Gamma}\langle e^{-i\Phi}\sigma^{+}_{j}+e^{i\Phi}\sigma^{-}_{j}\rangle+\xi(t).
\end{equation}
where $\Phi={\rm arg}\beta$ is the phase of the local oscillator, and $\xi(t)=dw(t)/dt$ represents Gaussian white noise satisfying normal distribution with $dw(t)$ a complex Wiener increment.
When the detector with efficiency $\eta$ is considered in the whole process, the above expression can be modified as
\begin{equation}\label{A4}
J^{\rm {hom}}_{j}(t)=\langle\sigma^{x}_{j}\rangle+\frac{\xi(t)}{\sqrt{\eta\Gamma}},
\end{equation}
where we have set $\Phi=0$ for simplicity.

\section{Steady-state solution of two-body $XXX$ spin-chain quantum battery}\label{B}
The Hilbert space of considered $XXX$ spin-chain quantum battery consists of the following four bases  $|1\rangle=\mid\uparrow\uparrow\rangle$, $|2\rangle=\mid\uparrow\downarrow\rangle$, $|3\rangle=\mid\downarrow\uparrow\rangle$, and $|4\rangle=\mid\downarrow\downarrow\rangle$. The density matrix $\rho$ of the system at an arbitrary time $t$ can be written as
\begin{equation}\label{JUZHEN}
\rho(t)=\left(
\begin{array}{cccc}
\rho_{11}(t)&\rho_{12}(t)&\rho_{13}(t)&\rho_{14}(t)\\
 \rho_{21}(t)&\rho_{22}(t)&\rho_{23}(t)&\rho_{24}(t)\\
 \rho_{31}(t)&\rho_{32}(t)&\rho_{33}(t)&\rho_{34}(t)\\
 \rho_{41}(t)&\rho_{42}(t)&\rho_{43}(t)&\rho_{44}(t)
\end{array}
\right).
\end{equation}
By solving Eq.~(\ref{M}), we obtain following coupled differential equation of diagonal elements
\begin{widetext}\label{C2}
\begin{equation}
\dot{\rho}_{11}(t)=-2\Gamma\rho_{11}(t)-4f\cos\alpha\rho_{11}(t)
+\frac{f^{2}}{\Gamma\eta}[\rho_{22}(t)+\rho_{33}(t)-2\rho_{11}(t)],
\end{equation}
\begin{equation}
\dot{\rho}_{22}(t)=2iJ[\rho_{23}(t)-\rho_{32}(t)]+(\Gamma+2f\cos\alpha)[\rho_{11}(t)-\rho_{22}(t)]
+\frac{f^{2}}{\Gamma\eta}[\rho_{11}(t)-2\rho_{22}(t)+\rho_{44}(t)],
\end{equation}
\begin{equation}
\dot{\rho}_{33}(t)=-2iJ[\rho_{23}(t)-\rho_{32}(t)]+(\Gamma+2f\cos\alpha)[\rho_{11}(t)-\rho_{33}(t)]
+\frac{f^{2}}{\Gamma\eta}[\rho_{11}(t)-2\rho_{33}(t)+\rho_{44}(t)],
\end{equation}
and $\dot{\rho}_{44}(t)=-\dot{\rho}_{11}(t)-\dot{\rho}_{22}(t)-\dot{\rho}_{33}(t)$. The corresponding differential equations of off-diagonal elements are
\begin{eqnarray}
\dot{\rho}_{12}(t)&=&\dot{\rho}_{21}^{\ast}(t)=-i\{h\rho_{12}(t)+2J[\rho_{12}(t)-\rho_{13}(t)]\}
-f\{[3\rho_{12}(t)+\rho_{21}(t)]\cos\alpha+i[\rho_{12}(t)+\rho_{21}(t)]\sin\alpha\}\nonumber\\&&-\frac{f^{2}}{\Gamma\eta}[2\rho_{12}(t)+e^{2i\alpha}\rho_{21}(t)-\rho_{34}(t)]-\frac{3\Gamma\rho_{12}(t)}{2},
\end{eqnarray}

\begin{eqnarray}
\dot{\rho}_{13}(t)&=&\dot{\rho}_{31}^{\ast}(t)=-i\{h\rho_{13}(t)+2J[\rho_{13}(t)-\rho_{12}(t)]\}
-f\{[3\rho_{13}(t)+\rho_{31}(t)]\cos\alpha+i[\rho_{13}(t)+\rho_{31}(t)]\sin\alpha\}\nonumber\\&&-\frac{f^{2}}{\Gamma\eta}[2\rho_{13}(t)+e^{2i\alpha}\rho_{31}(t)-\rho_{24}(t)]-\frac{3\Gamma\rho_{13}(t)}{2},
\end{eqnarray}
\begin{equation}
\dot{\rho}_{14}(t)=\dot{\rho}_{41}^{\ast}(t)=-e^{i\alpha}f[2\rho_{14}(t)+\rho_{23}(t)+\rho_{32}(t)]-2ih\rho_{14}(t)-\Gamma\rho_{14}(t)-\frac{f^{2}}{\Gamma\eta}\{2\rho_{14}(t)+e^{2i\alpha}[\rho_{23}(t)+\rho_{32}(t)]\},
\end{equation}
\begin{eqnarray}
\dot{\rho}_{23}(t)&=&\dot{\rho}_{32}^{\ast}(t)=2iJ[\rho_{22}(t)-\rho_{33}(t)]-\Gamma\rho_{23}(t)+\frac{f^{2}}{\Gamma\eta}\{-2\rho_{23}(t)-e^{-2i\alpha}[\rho_{14}(t)+e^{4i\alpha}\rho_{41}(t)]\}\nonumber\\&&
-f\cos\alpha[\rho_{14}(t)+2\rho_{23}(t)+\rho_{41}(t)]+if\sin\alpha[\rho_{14}(t)-\rho_{41}(t)],
\end{eqnarray}
\begin{eqnarray}
\dot{\rho}_{24}(t)&=&\dot{\rho}_{42}^{\ast}(t)=[\rho_{13}(t)-\frac{\rho_{24}(t)}{2}]\Gamma-i\{h\rho_{24}(t)+2J[\rho_{34}(t)-\rho_{24}(t)]\}
+\frac{f^{2}}{\Gamma\eta}[\rho_{13}(t)-2\rho_{24}(t)-e^{2i\alpha}\rho_{42}(t)]\nonumber\\&&+f\{[2\rho_{13}(t)-\rho_{24}(t)-\rho_{42}(t)]\cos\alpha-i[\rho_{24}(t)+\rho_{42}(t)]\sin\alpha\},
\end{eqnarray}
\begin{eqnarray}
\dot{\rho}_{34}(t)&=&\dot{\rho}_{43}^{\ast}(t)=[\rho_{12}(t)-\frac{\rho_{34}(t)}{2}]\Gamma-i\{h\rho_{34}(t)+2J[\rho_{24}(t)-\rho_{34}(t)]\}
+\frac{f^{2}}{\Gamma\eta}[\rho_{12}(t)-2\rho_{34}(t)-e^{2i\alpha}\rho_{43}(t)]\nonumber\\&&+f\{[2\rho_{12}(t)-\rho_{34}(t)-\rho_{43}(t)]\cos\alpha-i[\rho_{34}(t)+\rho_{43}(t)]\sin\alpha\}.
\end{eqnarray}
 Then by imposing the condition $\dot{\rho}(t)=0$, we
get the following steady-state solution
\begin{equation}\label{ww}
\rho_{11}(\infty)=\frac{f^{4}}{(2f^{2}+2f\Gamma\eta\cos{\alpha}+\Gamma^{2}\eta)^{2}}, \qquad \rho_{22}(\infty)=\frac{f^{2}(f^{2}+2f\Gamma\eta\cos{\alpha}+\Gamma^{2}\eta)}{(2f^{2}+2f\Gamma\eta\cos{\alpha}+\Gamma^{2}\eta)^{2}},
\end{equation}
and $\rho_{33}(\infty)=\rho_{22}(\infty)$, $\rho_{44}(\infty)=1-\rho_{11}(\infty)-\rho_{22}(\infty)-\rho_{33}(\infty)$.
According to Eq.~(\ref{energy}), the energy stored by the quantum battery is
\begin{eqnarray}\label{e1}
\Delta E_{XXX}(\infty)&=&(J+h)\rho_{11}(\infty)-J\rho_{22}(\infty)+2J\rho_{32}(\infty)+2J\rho_{23}(\infty)-J\rho_{33}(\infty)+(J-h)\rho_{44}(\infty)\nonumber\\&&
-[(J+h)\rho_{11}(0)-J\rho_{22}(0)+2J\rho_{32}(0)+2J\rho_{23}(0)-J\rho_{33}(0)+(J-h)\rho_{44}(0)].
\end{eqnarray}
\end{widetext}

\section{The highest energy state of a $N$-site $XXX$ spin chain}\label{Bbb}
For a general $N$-site $XXX$ spin chain ($\gamma=0$, $\Delta=1$), the corresponding Hamiltonian can be rewritten as

\begin{equation}\label{XXX}
H_{B}=h\sum\limits_{j=1}^{N}S_{j}^{z}+\sum\limits_{j=1}^{N-1}2J[(\boldsymbol {S}_{j}+\boldsymbol {S}_{j+1})^{2}-\boldsymbol {S}_{j}^{2}-\boldsymbol {S}_{j+1}^{2}],
\end{equation}
where $\boldsymbol {S}_{j}\equiv (S^{x}_{j}, S^{y}_{j}, S^{z}_{j})=(\sigma^{x}_{j}, \sigma^{y}_{j}, \sigma^{z}_{j})/2$. 
Since the two parts of above Hamiltonian commute, it's easy to check that highest-energy eigenstate of the $XXX$ spin chain is $\mid\uparrow\uparrow\cdots\uparrow\rangle$ for all positive real parameters, which is independent of the nearest-neighbor coupling strength between spins, and the corresponding eigenenergy is $E_{\rm max}=Nh/2+(N-1)J$.

\begin{figure}
\centering\scalebox{0.21}{\includegraphics{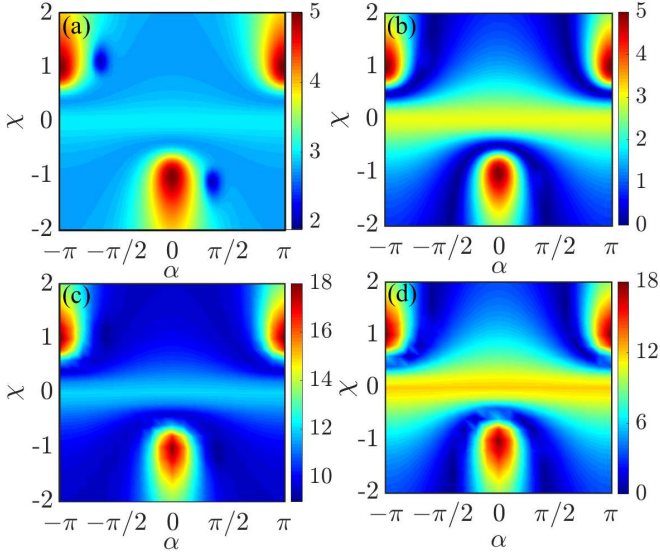}}
\caption{\label{MM} (a)-(b) describe the stored energy and ergotropy of the battery with spin number of $N=2$ as functions of $\alpha$ and $\chi$, respectively. Other parameters are $\eta=1$ and $J=h$. (c)-(d) are the same as (a)-(b), but the spin number is $N=6$.}
\end{figure}

\begin{figure}
\centering\scalebox{0.4}{\includegraphics{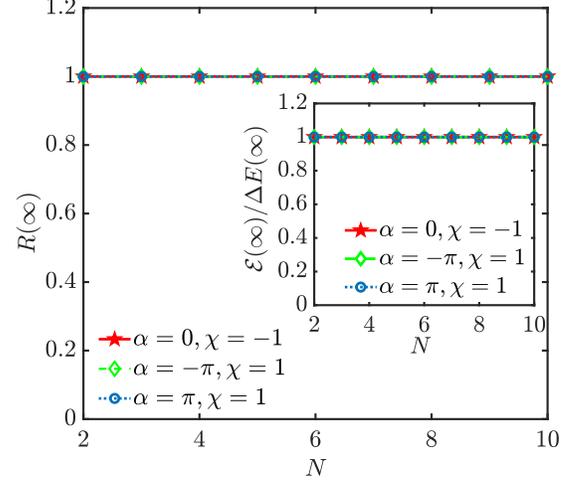}}
\caption{\label{VV} The value of $R(\infty)$ and $\mathcal{E}(\infty)/\Delta E(\infty)$  of the battery with different spin number under the three groups of optimal feedback parameters, where the measurement efficiency is $\eta=1$.
 }
\end{figure}

In the Sec.~\ref{IIIB}, we have observed that the optimal feedback conditions of the quantum battery composed of four spins are the same as that of the two spins. Now we present the numerical results of the energy storage and ergotropy of the battery composed of six spins varying with parameters $\alpha$ and $\chi$ in Fig.~\ref{MM}. For convenience, we also reproduce the numerical simulation results of the quantum battery composed of two spins. 
Fig.~\ref{VV} further depicts the variation of effective space utilization rate and $\mathcal{E}(\infty)/\Delta E(\infty)$ in the steady state with the number of spin, under the three groups of optimal feedback conditions. These results can reflect to some extent that our feedback scheme is independent of the number of spins constituting the spin-chain quantum battery.
\begin{figure}
\centering\scalebox{0.23}{\includegraphics{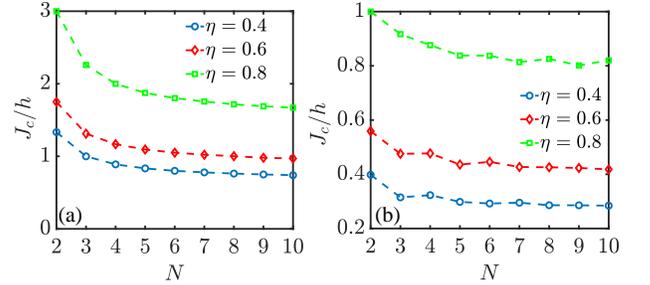}}
\caption{\label{P10} (a) The critical value $J_{c}$ of spin-spin interaction in energy storage varies with the spin number. (b) The critical value $J_{c}$ of spin-spin interaction in energy extraction (ergotropy) varies with the spin number. Different curves correspond to different measuring efficiency $\eta$.}
\end{figure}

\section{The charging process under imperfect measurement}\label{C}
Under imperfect measurement conditions ($\eta<1$), if the ground state of the battery with $N=2$ is $\mid\downarrow\downarrow\rangle$ ($J<h/4$), the stored energy of the battery in the steady state is
 \begin{equation}\label{EXXXxia}
\Delta E_{1}(\infty)=\frac{2(h-2J)\vartheta(\eta+2\chi\eta\cos\alpha)}{2\chi^{2}
+2\chi\eta\cos\alpha+\eta}+4(h-J)\vartheta^{2},
\end{equation}
where $\vartheta=\chi^{2}/(2\chi^{2}+2\chi\eta\cos\alpha+\eta)$. Since $(h-2J)>0$ at this time, $\Delta E_{1}(\infty)$ can always be maximized by selecting $\chi=1$, which means the feedback control is still valid under the condition of imperfect measurement.
While if the ground state of the system is $(\mid\uparrow\downarrow\rangle-\mid\downarrow\uparrow\rangle)/\sqrt{2}$ ($J>h/4$), the corresponding stored energy of the battery in the steady state becomes
\begin{equation}\label{EXXX}
\Delta E_{2}(\infty)=-h+4J+4J\vartheta^{2}+2(h-2J)\vartheta.
\end{equation}
The third and fourth terms in the r.h.s. of Eq.~(\ref{EXXX}) are mutually restricted for a large $J$, thus the feedback control may fail when $J$ exceeds a certain value $J_{c}$ which can be   
determined as $J_{c}/h=(2-\eta)/2(1-\eta)$ according to $\Delta E_{2}(\infty)\mid_{(\chi)=1}=\Delta E_{2}(\infty)\mid_{\chi=0}$.

In order to more intuitively reflect the influence of measurement efficiency $(\eta)$ and spin number ($N$) on the critical coupling strength $J_{c}$, we describe the variation of the critical value $J_{c}$ with the spin number under different measurement efficiencies from the perspective of energy storage and extraction in Figs.~\ref{P10} (a) and \ref{P10} (b), respectively.

\bibliography{yao}
\end{document}